\documentclass[pre,aps,twocolumn]{revtex4-1}
\usepackage{graphicx}
\usepackage{bm}
\usepackage{tabularx}
\usepackage{amsmath}

\begin{document}

\title{Driven coupled Morse oscillators --- visualizing the phase space and characterizing the transport}

\author{Astha Sethi}
\altaffiliation{Current address: Department of Chemistry, Chemical Physics Theory Group, and Center for quantum Information and Quantum Control, University of Toronto, M5S 3H6 Toronto, Canada}
\affiliation{Department of Chemistry, Indian Institute of Technology, Kanpur, Uttar Pradesh 208016, India}
\author{Srihari Keshavamurthy}
\email{srihari@iitk.ac.in}
\affiliation{Department of Chemistry, Indian Institute of Technology, Kanpur, Uttar Pradesh 208016, India}

\begin{abstract}
Recent experimental and theoretical studies indicate that intramolecular energy redistribution (IVR) is nonstatistical on intermediate timescales even in fairly large molecules. Therefore, it is interesting to revisit the the old topic of IVR versus quantum control and one expects that a classical-quantum perspective is appropriate to gain valuable insights into the issue.  However, understanding classical phase space transport in driven systems is a prerequisite for such a correspondence based approach and is a challenging task for systems with more then two degrees of freedom.
In this work we undertake a detailed study of the classical dynamics of a minimal model system - two kinetically coupled coupled Morse oscillators in the presence of a monochromatic laser field.  Using the technique of wavelet transforms a representation of the high dimensional phase space, the resonance network or Arnold web, is constructed and analysed. The key structures in phase space which regulate the dissociation dynamics are identified. Furthermore, we show that the web is nonuniform with the classical dynamics exhibiting extensive stickiness, resulting in anomalous transport. Our work also shows that pairwise irrational barriers might be crucial even in higher dimensional systems.  
\end{abstract}

\maketitle
\section{Introduction}

The influence of intramolecular vibrational energy redistribution (IVR) on the control of quantum dynamics has been a subject
of some debate ever since the first lasers gave hope to the dream of doing mode-specific chemistry\cite{al77,l77,by78,z80,jlr81}. In fact,  arguing
that the energy flow process is typically faster than the reaction timescales\cite{or79} and essentially statistical, researchers identified
IVR as the main culprit which spoils the idea of mode-specific control\cite{zb88}. 
Consequently, most of the quantum control methods shifted their focus away from explicitly controlling or beating the IVR process. Instead, the various control strategies
\cite{bsbook,gr97,rz00} concentrated on manipulating the quantum interferences, for which a prior knowledge of the field-free IVR mechanism is supposedly not required.
 
The observations above, however,  need to be reappraised in 
light of the recent  developments in our understanding of IVR. Both experimental\cite{exivr1,exivr2,exivr3} 
and theoretical\cite{thivr1,thivr2,thivr3} studies
indicate that IVR at intermediate times is nonstatistical even in 
moderately sized molecules. Specifically, from a molecular state-space perspective\cite{stsp0,stsp1,stsp2}, 
IVR is anisotropic with  local, as opposed to global,
density of states playing a crucial role and significant deviations from the Fermi golden rule regime in the form of power law
decays are to be expected\cite{sw93,mg98}. 
These studies certainly give rise to new hopes for controlling the IVR process itself and there has been 
some promising work\cite{ivrcontrol1,ivrcontrol2,ivrcontrol3} in this direction. However, 
in light of the recent advances, there is also a case to be made for  reopening the old debate -  can mechanistic insights into the energy flow dynamics  lead to better control 
strategies? It is not
entirely unreasonable to expect that a subtle  interplay between the IVR dynamics and the controlling field must occur in many of the quantum control studies\cite{onfzr2001,bgm2011}.
Hence, is it possible to envisage situations wherein IVR might actually aid, rather than thwart,
the control process?   An interesting ``inverse" question can be asked at this stage. What features of a control field, as obtained through 
optimal\cite{swr88} control theory, are a direct consequence of the IVR dynamics?  There is a striking parallel here to the more contemporary issue of the search for decoherence-free subspaces in quantum control of open systems\cite{dfs}. A mechanistic understanding of the interplay between decoherence and control of a system is proving to be invaluable in a wide range of areas. Therefore, although answering the inverse question posed above is far from easy, one hopes that a careful study of model systems can lead to a finer understanding of the role of IVR in quantum control of molecular processes.

In order to uncover the mechanism and influence of IVR on quantum control one can adopt a purely quantum, purely classical, or a semiclassical approach. However, there is little doubt that detailed insights come from studying and comparing all of the approaches.  In particular, although  quantum studies of fairly high dimensional systems are now possible, 
a key reason for taking the classical-quantum correspondence perspective has to do with the fact that classical dynamics provides the correct baseline to gauge the extent to which quantum effects like interference and tunneling are important for control. Such a classical-quantum correspondence perspective has proved useful not only in the context of IVR and control but also
in many other areas like driven atomic systems\cite{drivatom}, mesoscopic systems\cite{mesoqc} and trapped cold atoms\cite{bec}. 
Various studies on systems with two or fewer degrees of freedom ($N \leq 2$) have led to the identification of key phase space structures that regulate IVR and, more importantly, their quantum fingerprints. Thus, the role of chaos, Kolmogorov-Arnold-Moser (KAM) barriers, partial barriers like broken separatrices\cite{cl80} and cantori\cite{cantorirefs}, 
and phase space bifurcations\cite{psbifur} have been extensively studied\cite{ivr2drev1,ivr2drev2}. 

Nevertheless, there are two major obstacles in extending the correspondence approach to systems with more than two degrees of freedom. The first obstacle is the difficulty in visualizing the global dynamics in high dimensional phase spaces. The second obstacle has to do with the fact that the nature of phase space transport changes dramatically in going from two to three degrees of freedom. For example, KAM tori are no longer true barriers in the phase space and there are no obvious generalizations of the notion of cantori in $N \geq 3$. In addition, Hamiltonian systems typically exhibit ``stickiness"\cite{stickyall}  
{\it i.e.,} classically chaotic trajectories spending significant amount of time near regular structures or their remnants in the classical phase space.
Stickiness leads to long time dynamical correlations and is key to understanding the nature of phase space transport. Important insights into the universality of stickiness in Hamiltonian systems are starting to emerge\cite{sticky0,sticky1,sticky2,sticky3,sticky4,sticky5,sticky6} but their quantum manifestations, if any, are mostly unexplored. These obstacles, technical as well as conceptual, have been largely responsible for a waning of interest among researchers in studying and utilizing the insights afforded by a classical-quantum correspondence approach.  

In the last few years, however, there has been a rapid progress in our ability to investigate phase space transport in such high dimensional systems. Partly, this has to do with the recent exciting work\cite{tsrev1,tsrev2,tsrev3} 
in transition state theory - construction of locally recrossing free dividing surfaces
in phase space and dynamics near saddles on potential surfaces for $N \geq 3$ systems.  In turn, such studies have led to 
researchers extending the concept of a transition state for driven systems\cite{kbjbpu07} with obvious relevance to quantum control. On the other hand, attempts are also being made\cite{webref1,webref2,webref3} to connect the nature of global phase space transport with the structure of the nonlinear resonance networks, also known as the Arnold web. In particular, the Arnold web is central for a classical description of the IVR dynamics. Although a precise correspondence between the classical  dynamics  on the Arnold web and the quantum  dynamics in the molecular state space is lacking, there is sufficient reason to expect that the classical insights will prove crucial in unearthing the mechanisms of IVR in a molecule. Will the classical insights also help in controlling the intramolecular energy flow itself or quantum control of processes influenced by IVR? At present this question remains unanswered. These recent advances combined with some of the early pioneering work collectively motivate the present study, which is a first step in answering the question posed above.

In this work we take a minimal model of two kinetically coupled Morse oscillators, describing two local modes, driven by a monochromatic field to understand the interplay of IVR and control in molecular dissociation dynamics. More specifically, we consider parameters appropriate for the
HCN (hydrogen cyanide) molecule. HCN has been a benchmark ``small'' molecular system
for several quantum control studies\cite{cb91,bl04,hobi02,brr95} whose primary, and rather challenging,  aim is to dissociate  the stronger C$\equiv$N bond while 
keeping the weaker C$-$H mode ``quiet''. However, in the current study we do not attempt the selective breaking of the C$\equiv$N bond and instead study the C$-$H bond dissociation. The focus here, in this $N=2.5$ degrees of freedom system, is on identifying the main classical phase space structures that regulate the dissociation dynamics. In Sec.~\ref{modham} we give a brief description of the model Hamiltonian. The details of the construction of the Arnold web are discussed in Sec.~\ref{arweb} and the important features in phase space are identified. Interestingly, it is observed that the crucial regions in the web correspond to classical phase space structures with highly irrational frequency ratios. These results are reminiscent of the early work by Martens, Davis, and Ezra\cite{mde87}
wherein the importance of pairwise noble barriers to IVR in the OCS molecule ($N = 3$) was highlighted. In Sec.~\ref{hubsrole} we study the role of two important regions (``hubs") in the resonance network.  One of them, the ``dissociation hub", acts as a gateway for dissociating trajectories and possibly of some interest in the context of transition state theory. In the other hub, the ``noble hub", we establish the extensive stickiness of the dynamics with significant implications for transport in the phase space. Section~\ref{conclusions} concludes.

\section{Preliminaries: model Hamiltonian and motivation}
\label{modham}
 
We consider two kinetically coupled Morse oscillators being driven by a monochromatic field.
The corresponding classical Hamiltonian, considered in an earlier work\cite{bl04} as well, is given by
\begin{equation}
H=H_{0}+H_{I} \equiv H_{CH} + H_{CN} + T_{\rm coup} + H_{I}
\label{copmorse}
\end{equation}
where the matter part of the Hamiltonian is
\begin{subequations}\label{full_ham}
\begin{eqnarray}
H_{CH}&=&\frac{p_{x}^2}{2M_{CH}}+D_{CH}\left(1-e^{-\alpha x}\right)^{2} \label{eq_CH}\\
H_{CN}&=&\frac{p_{y}^2}{2M_{CN}}+D_{CN}\left(1-e^{-\zeta y}\right)^{2} \label{eq_CN}\\
T_{\rm coup}&=&-\frac{p_{x}p_{y}}{M_{C}}\label{cop}.
\end{eqnarray}
\end{subequations}
and the matter-radiation interaction in the dipole limit is taken to be
\begin{equation}
H_{I}=-\lambda_{F}\mu(x)\cos(\omega_{F} t).
\end{equation}
The form of the dipole function\cite{bl04} is given by the expression
\begin{equation}
\mu(x)=e^{-\eta (x+x_{e})}\sum_{i=1}^{4}A_{i}(x+x_{e})^{i}
\label{chdip}
\end{equation}
and the values of the various parameters can be found in the earlier work\cite{bl04}.

As indicated In Eqn.~\ref{copmorse}, the various parameters of $H_{0}$ are appropriate for the HCN molecule. 
Thus, $H_{CH}$ and $H_{CN}$ are Morse oscillator 
Hamiltonians for the anharmonic vibrational displacements of the C$-$H ($x$) and C$\equiv$N ($y$) bonds respectively with $x_{e}$ being the
equilibrium bond length of the CH mode.
The reduced masses of the CH and the CN modes are denoted by $M_{CH}$ and $M_{CN}$ respectively and $M_{C}$ denotes the
mass of the carbon atom. Note that the number of bound states of the CH and the CN oscillators are about $N_{b}=27$ and $N_{b}=80$ respectively for the given parameters. 
The vibrational local modes are directly coupled to each other via the kinetic coupling $T_{\rm coup}$ and is 
responsible for IVR in the system. In addition, an indirect coupling is also present between the two modes due to the interaction with
the laser field via $H_{I}$.

Throughout this work the various parameters corresponding to the Morse oscillators and the driving  field
are all in atomic units. The driving field strength $\lambda_{F}=0.009$ ($\sim 3$ TW/cm$^{2}$)
and frequency $\omega_{F} = 0.011$ ($\sim 0.5$ fs$^{-1}$)
are chosen to avoid any significant excitation of CN bond, since we are 
interested in studying the effect of driving field only on the CH mode.

A few comments on the choice of the model Hamiltonian and the form of the matter-field coupling is appropriate at this stage. Firstly, the form of $H_{0}$ has a long history in chemical physics starting with the beautiful work by Thiele and Wilson\cite{tw61} and the extensive studies by Oxtoby and Rice\cite{oxrice76} on the connection between nonlinear resonance overlaps and IVR. 
This choice is simple enough to use the analytically known action-angle variables for a Morse oscillator but has the basic features of interest to us - IVR, bifurcations, and a mixed regular-chaotic phase space. Secondly,
as mentioned in the introduction, our goal here is to mechanistically
understand the role of IVR in regulating the dissociation dynamics and identify the key phase space structures.
In this initial study, therefore, we choose to avoid complications from other competing processes that may hinder a clear interpretation.
Thus, we choose a simple monochromatic form for the driving field
and it is important to note that this model already has a five-dimensional phase space. For the same reason we choose to couple only the CH mode to the field. For the weak driving fields (the ionization threshold for HCN is around $100$ TW/cm$^{2}$) and the specific initial states considered here (see next section) this is reasonable. Moreover, even upon coupling the CN mode to the field, our study indicates no significant change in the qualitative nature of the phase space structures discussed later.
This minimal model, therefore, has the basic ingredients {\it i.e.,} IVR , a field attempting to dissociate a bond, and an essential degree of dynamical complexity. In addition, such a model represents the next level of difficulty in extending the insights gained from studies on one dimensional driven Morse oscillator system\cite{gogmil88} to systems with higher degrees of freedom. Note that the parameters are representative of HCN but the model is still a reduced dimensional one since the bending mode is not taken into account. It is known from a recent work\cite{gmr05}
that the bend mode is important to the dissociation dynamics. However, understanding the phase space transport in the current reduced dimensional model is a prerequisite for attempting the more challenging case of including all three modes of the molecule.

 \begin{figure*}[t]
\includegraphics[width=0.9\textwidth]{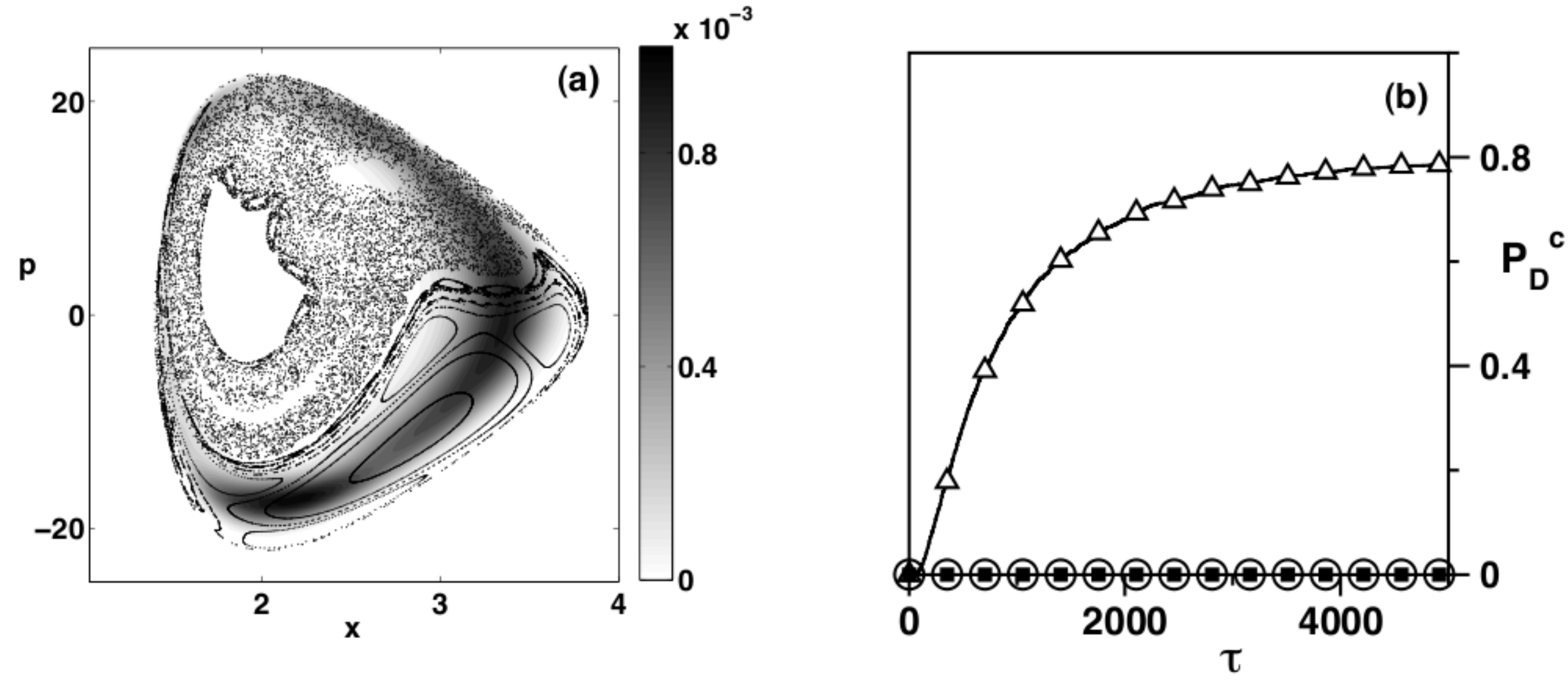}
\caption{(a) Husimi distribution of the initial state $|8,3\rangle$ superimposed
on the corresponding CH-mode Poincar\'{e} surface of section in the absence of the driving field.
(b) The classical dissociation probability $P_{D}^{c}$ for the CH bond in the presence of the driving field
(triangles) with strength $\lambda_{F}=0.009$ and frequency $\omega_{F}=0.011$.
Note that the CH bond does not dissociate until 5000$T_{F}$ in the absence of
the driving field (squares, IVR alone) or 
when the kinetic coupling between CH and CN mode is
switched off (circles, field alone).}
\label{fig1}
\end{figure*} 

\subsection{Motivation}
\label{motive}

In order to motivate the current study and connect to the discussion in the introduction we start
by investigating the dissociation dynamics of the
zeroth order state $|n_{CH},n_{CN}\rangle=|8,3\rangle$.
This specific choice of state has to do with the 
fact that in the absence of laser field the classical 
phase space exhibits several interesting structures at the corresponding zeroth-order energy. 
Indeed, the Poincar\'{e} surface of section shown in Fig.~\ref{fig1}(a) exhibits mixed
regular-chaotic regions with prominent $\Omega_{CH}:\Omega_{CN}=1$:$1$ 
and $\Omega_{CH}:\Omega_{CN}=3$:$2$ nonlinear resonances interspersed in the chaotic sea. In addition,
the Husimi distribution\cite{H40} of $|8,3\rangle$ superimposed on the
surface of section in Fig.~\ref{fig1}(a) shows that the state is mostly localized
in the $\Omega_{CH}:\Omega_{CN}=1$:$1$ resonance region in the phase space. Note that $|8,3 \rangle$ is
not an eigenstate of the  matter Hamiltonian $H_{0}$ and thus the Husimi distribution does spread across the border between the main 
1$:$1 island and the chaotic sea.  For this typical $N=2$ KAM system, the regular-chaos border has a hierarchical structure which is expected to result in sticky dynamics. In particular, a very high order resonance island is visible in the border region in Fig.~\ref{fig1}(a) which suggests the existence of cantori in the phase space. Such barriers, partial or otherwise, limit the extent of classical IVR in the system.  Specifically, in the undriven subsystem $(\lambda_{F}=0)$, an analysis along the lines of the earlier work\cite{oxrice76} by Oxtoby and Rice reveals that  the overlap of low order resonance zones is not extensive enough at the energy of interest to result in dissociation. 
Thus, IVR alone is incapable of  dissociating the CH bond. This is confirmed in Fig.~\ref{fig1}(b) where 
the classical dissociation probability is zero for the entire time duration of interest.
We stress at this juncture that any other initial state is equally a good choice as long as the generic criteria of a mixed phase space and partially localized Husimi distribution are satisfied.
For instance other states such as $|0,15\rangle$ and $|4,8\rangle$ yield similar qualitative insights as the one chosen here. 

We now introduce the driving field with parameters mentioned above and note that
$\omega_{F}=0.011$ is chosen so as to be off-resonant with respect to the zeroth order state of interest.
To begin, we decouple the CH and CN modes by turning off the kinetic coupling term $T_{\rm coup}$ in Eq.~\ref{cop} and drive the CH mode. 
Classical trajectories are propagated by integrating the Hamilton's equations of motion up to a final time 
 $5000 T_{F}$ with $T_{F} \equiv 2\pi/\omega_{F} \sim 14$ fs being the field period. Trajectories are  considered\cite{mw82} to be dissociated  if the displacement of the bond of interest exceeds a certain threshold, set here to be $15$ a.u. for
the CH bond dissociation.  For the chosen driving field strength it is clear from Fig.~\ref{fig1}(b) that again there is negligible dissociation of the CH bond\cite{chiricomment}. Hence, the mode-mode coupling {\it i.e.,} IVR in the absence of the field or the field in the absence of IVR are incapable of dissociating the CH bond. However, and this is the key observation, the weak driving field with nonzero mode-mode coupling does lead to large dissociation probability, as seen in Fig.~\ref{fig1}(b) (triangles) and poses an interesting question. What is the precise mechanism,  undoubtedly involving a subtle interplay between IVR and the driving field, by which the initial state $|8,3 \rangle$ undergoes dissociation? In the following section we construct the Arnold web\cite{arwebcomment} and attempt to answer the question from the perspective of transport in the classical phase space.

\section{Time-frequency analysis and visualizing the phase space}
\label{arweb}

The  example summarized in Fig.~\ref{fig1} motivates us to understand the phase space
transport in the system of interest.
However, the phase space in this case is five dimensional and not easily visualized.
On the other hand, recent studies on IVR from time dependent
perspective\cite{mde87,webref2} have established the importance of the resonance network or Arnold web {\it i.e.,}
the various resonances and their disposition in the phase space. Therefore, following the dynamics of an initial state
on the resonance network should reveal the dominant frequency
lockings experienced by the state  as a function of time. Consequently,
one can obtain valuable information regarding the local structures in the phase 
space which regulate the dissociation dynamics in the presence of laser field. Our interest here is to test if such an approach would indeed provide the mechanistic insights.  

To begin with one needs to construct  the Arnold web and then interpret the dynamics of the system. The implementation aspect has been dealt with by several studies over the last decade or so. Starting with the early pioneering work by Martens, Davis, and Ezra\cite{mde87} on the use of windowed Fourier-transforms and a little later by Laskar\cite{l93}, several methods like fast Lyapunov indicators\cite{ftl},
mean exponential growth of nearby orbits\cite{megno}, and frequency modulation indicator\cite{fmi} have been proposed. However, the interpretation of the classical dynamics is not so straightforward for nonintegrable systems for reasons mentioned in the introduction. In the next subsection we utilize a wavelet based technique, proposed by Arevalo and Wiggins\cite{aw01}, to construct  the resonance network for the system of interest. Subsequently, we attempt to provide an interpretation of the dynamics based on the network and highlight the role of IVR in presence of an external driving field.

\subsection{Constructing the resonance network}

\begin{center}
\begin{figure}[t]
\includegraphics[width=0.5\textwidth]{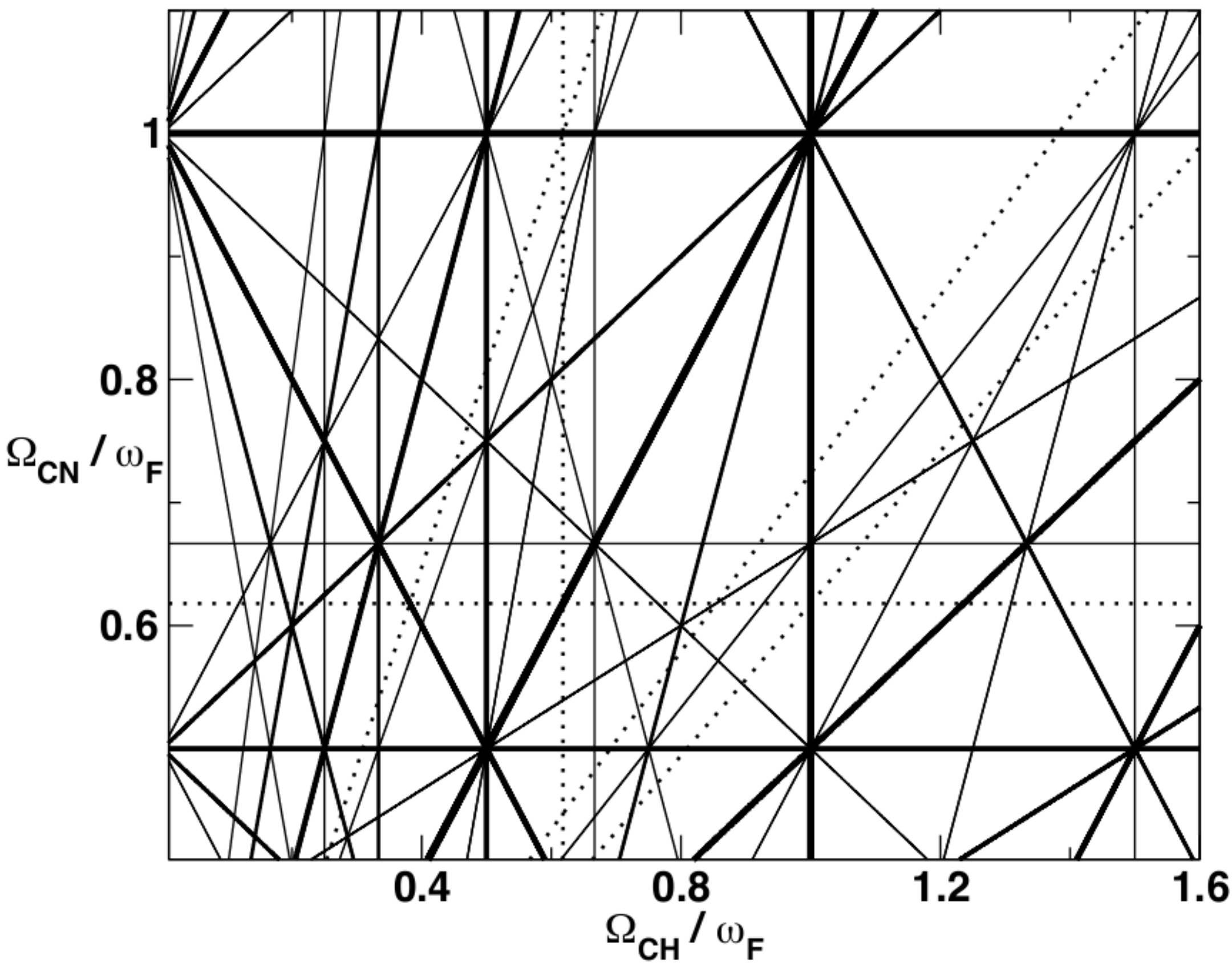}
\caption{Arnold web for the system of interest showing all the resonances of order $\sum_{} |k_{j}| \leq 5$, satisfying the condition in Eq.~\ref{res_cond},  as a frequency ratio space. As a guide the lines are shown with thickness inversely proportional to the order of the resonance. Vertical and horizontal lines correspond to mode-field resonance whereas positively sloped lines correspond to mode-mode resonance. The dotted lines show some of the possible pairwise noble barriers. See text for discussions.}
\label{fig1p}
\end{figure}
\end{center}

Since the exact action-angle variables $(J,\theta)$ for a Morse oscillator are known one can explicitly write down the 
zeroth-order nonlinear frequencies. Thus, for the CH oscillator one has
\begin{equation}
\Omega^{(0)}_{CH}(J) = \omega_{CH} \left[1 - \frac{\omega_{CH}}{2D_{CH}} J \right]
\end{equation}
with $\omega_{CH}$ being the harmonic frequency of the CH mode. A similar expression for the CN oscillator can be written down.
The general  resonance  condition involving the two oscillators and the driving field frequency can be expressed as
\begin{equation}
k_{1} \Omega^{(0)}_{CH}+k_{2}\Omega^{(0)}_{CN}+k_{3} \omega_{F}=0 
\label{res_cond}
\end{equation}
with $(k_{1},k_{2},k_{3})$ being a set of mutually prime integers. The order of a given resonance is $\sum_{j} |k_{j}|$
and the set of all resonances up to a given order 
forms the Arnold web at that order. Figure~\ref{fig1p} shows the web of resonances up to a total order of five. Pure mode-mode (binary, $k_{3}=0$) resonances appear on the web as lines with positive slopes and do
not involve the field. In contrast, ternary resonances ($k_{1},k_{2} > 0$ and $k_{3} < 0$) involve both the modes and the field and appear as lines with negative slopes. Intersection of the binary and ternary resonances result in mode-field resonances (vertical and horizontal lines) and such intersections form resonance junctions. Some of the resonance junctions, for example at $(1,1)$, $(1/2,1)$, and $(3/2,1/2)$ in Fig.~\ref{fig1p} are clearly visible. The resonance junctions are expected to be important for the phase space transport.
However, such a ``static'' construction cannot {\it a priori} highlight the dynamically
significant regions of the web except in the limit of small perturbations\cite{l93,aw01}. 
The system of interest here is certainly far  from near-integrability (cf. Fig.~\ref{fig1}(a)) and hence the static web can only act as a guide. In particular, irregular trajectories can have several frequency components and even the dominant among them evolve non trivially with time.
It is hence necessary to construct the ``dynamical" web from the actual nonlinear
frequencies $(\Omega_{CH}(t),\Omega_{CN}(t))$ determined as a function of time.

Classical trajectories are computed over a time interval $[0,T]$ starting with
the zeroth-order actions corresponding to the initial state of interest.
The dynamical phase space function $z_{l}(t)=x_{l}(t)+i p_{l}(t)$ corresponding to the $l^{\rm th}$ mode is
subjected to a continuous wavelet transform\cite{aw01}
\begin{equation}
L_{\psi}z_{l}(a,b) = a^{-1/2} \int_{-\infty}^{\infty} z_{l}(t)
\psi^{*}\left(\frac{t-b}{a}\right) dt,
\label{wavelet}
\end{equation}
with $a > 0$ and real $b$. As in the previous studies\cite{aw01,webref2,webref3} we choose
a Morlet-Grossman form for the mother wavelet
\begin{equation}
\psi(t) = \frac{1}{\sigma \sqrt{2 \pi}} e^{2 \pi i \lambda t}
e^{-t^{2}/2 \sigma^{2}},
\end{equation}
with $\lambda=1$ and $\sigma=2$.
The wavelet transform in Eq.~\ref{wavelet} yields the frequency content
of $z_{l}(t)$ over a time window around $t=b$. Typically, for an integrable or regular motion there is a single dominant frequency characterizing the phase space structure over the entire time. However, an irregular trajectory is associated with several frequency components.
In principle it is important to take into consideration all the frequency components of
the time evolved classical trajectories. However, here we work with the dominant frequency at a given time which is computed
by determining the scale ($a$, inversely proportional to frequency)
which maximizes the modulus of the wavelet transform, {\it i.e.,}
$\Omega_{l}(t=b)={max}_{a}|L_{\psi}z_{l}(a,b)|$.
This yields the required nonlinear frequencies $\Omega (t)$ and the dynamics
is followed in the frequency ratio space\cite{mde87} $(\Omega_{CH}/\Omega_{F},\Omega_{CN}/\Omega_{F})$.
Note that previous works suggest that
the web generated using the dominant frequencies is
still capable of providing important information
on the nature of the classical dynamics\cite{webref2,webref3}.

The frequency ratio space is divided into cells and the number of times 
that a cell is visited by all the trajectories is presented as a density plot. We further normalize
the highest density to one for convenience. 
For the sake of clarity the resulting network is shown on a logarithmic scale in order to highlight the significant regions.

\subsection{Interpretation of the dynamical resonance network}
\label{arwebinterp}

\begin{figure}[htbp]
\begin{center}
\includegraphics[width=0.55\textwidth]{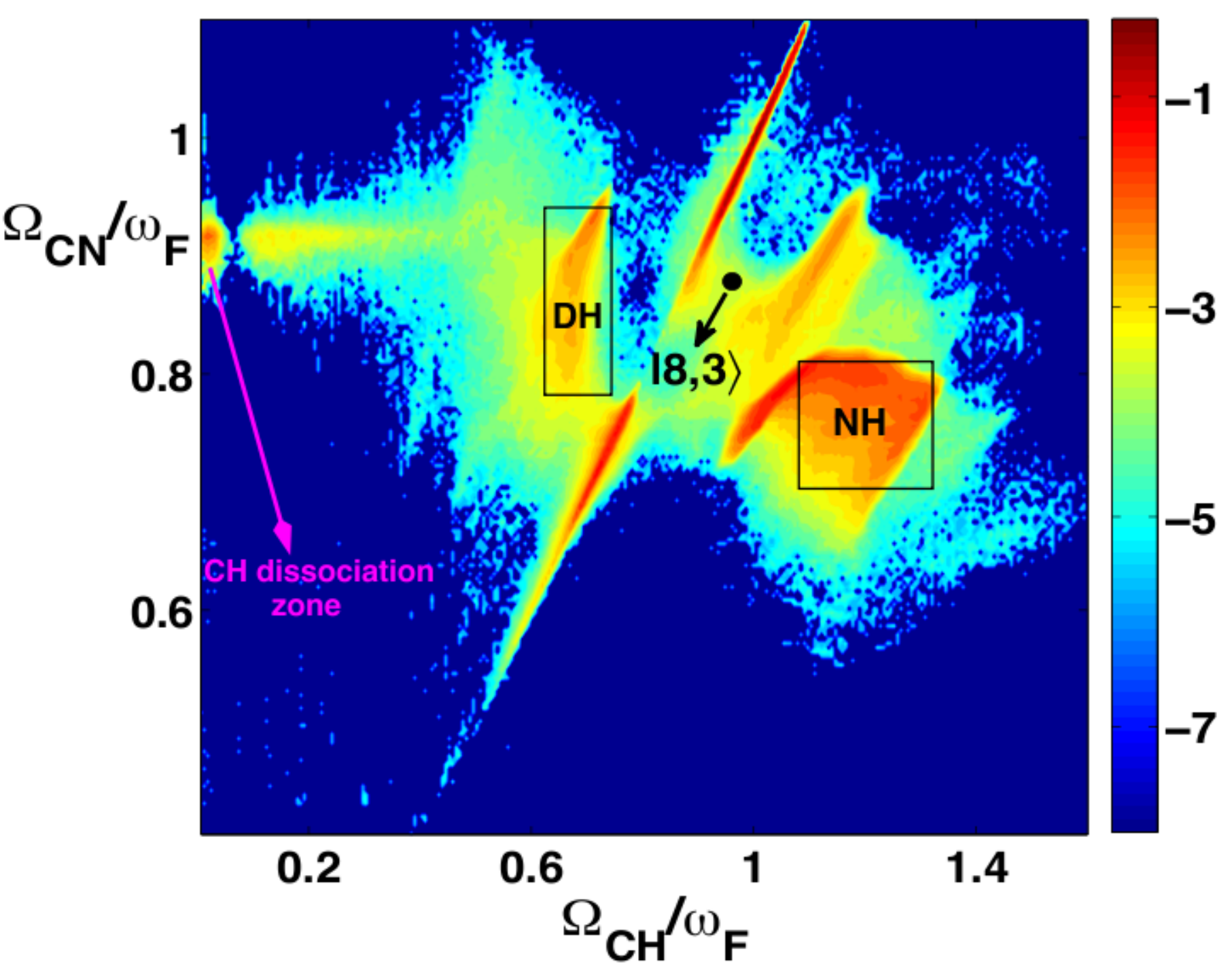}
\caption{Dynamical frequency ratio space (Arnold web)  corresponding to the initial state
$|n_{CH},n_{CN} \rangle=|8,3\rangle$, constructed via time-frequency analysis of
an ensemble of $5000$ trajectories propagated from $\tau=0$ to $\tau=5000$ $T_{F}$.
Note that the normalized density is plotted on
the logarithmic scale in order to highlight the sparse yet key regions in the phase space.
Two key regions denoted DH and NH (enclosed by squares),  the location of the initial state (black circle), and
the CH dissociation region are also indicated. }
\label{fig2}
\end{center}
\end{figure}

The resulting density plot for the state $|8,3\rangle$ is shown
in Fig.~\ref{fig2} in the frequency ratio space $(\Omega_{CH}/\omega_{F},\Omega_{CN}/\omega_{F})$. Comparing this dynamical frequency ratio space to the static construction shown in Fig.~\ref{fig1p} reveals several crucial aspects of the dissociation dynamics.
First and foremost,  the heterogeneous nature of the dynamical frequency ratio space  indicates that the phase space dynamics is nonuniform. Thus,
there exist dynamically significant regions in the phase space which, based on the earlier work\cite{webref1} by Shojiguchi {\it et al.} on a different model system, are expected to strongly influence the dissociation dynamics. It is worth mentioning that several regions of the static web are not dynamically relevant for the specific initial state and timescale of interest. Similar observations have been made\cite{l93} earlier in the context of coupled standard maps.
Secondly, a relatively narrow region of the frequency ratio space around
$\Omega_{CN}/\omega_{F} \approx 0.9$ corresponds to the dissociation of the CH mode (characterized by low values of $\Omega_{CH}/\omega_{F}$). 
Thirdly, a careful look at the web in Fig.~\ref{fig2} indicates the presence of
three distinct regions in the phase space and below we focus on the nature of these regions.

The first prominent region in Fig.~\ref{fig2} corresponds to the  
$\Omega_{CH}:\Omega_{CN}=1:1$ nonlinear resonance. Note that the Husimi distribution of $|8,3\rangle$ in Fig.~\ref{fig1}(a) does exhibit substantial localization in the large $1$:$1$ island. However, the enhanced density observed in Fig.~\ref{fig2} is nontrivial since the invariant tori associated with the $1$:$1$ resonance are no longer true barriers in presence of the driving. As seen later, the classical trajectories do leave the resonance zone but repeatedly get captured into the resonance with large residence times. Consequently, the enhanced density near the $1$:$1$ resonance has serious implications for the nature of the phase space transport.
A second important region of the frequency ratio space 
is marked as ``NH'' (noble hub)  in Fig.~\ref{fig2} with most 
of the activity occurring around a hub consisting
of several mode-mode irrational (noble) frequency ratios and high-order resonances. 
The noble numbers are  highly irrational numbers of the form $(n+n'\gamma_{g})/(m+m'\gamma_{g})$ with $nm'-mn'=\pm1$ and $\gamma_{g}^{-1} \equiv (\sqrt{5}+1)/2$ being the golden ratio.
High density around $\Omega_{CH}=\Omega_{CN}$ resonance and
the noble hub indicates continuous
interaction between CH and CN modes via IVR that
seems to regulate the dissociation dynamics of the CH bond
in the presence of the driving laser field. 
An equally important region of the frequency ratio space,
marked as ``DH'' (dissociation hub) in Fig.~\ref{fig2}, is in the vicinity of the $\Omega_{CH}/\omega_{F}=\gamma_{g}-1$
irrational ratio. Interestingly, the narrow dissociation zone mentioned above lies mainly to the left of this DH region.

The enhanced density at the NH and DH hubs suggests an intriguing possibility - regions with pairwise noble frequency ratios seem to be acting as barriers to transport in higher degrees of freedom systems.  At this stage, it is
worth recalling an earlier classic study by Martens, Davis and Ezra\cite{mde87} on the OCS molecule wherein it
was conjectured that pairwise noble barriers in phase space might play a crucial role
in the dynamics of systems with more than two degrees of freedom. A recent detailed study\cite{pcu2009} by Paskauskas, Chandre, and Uzer clearly establishes that two-dimensional invariant tori can act as phase space bottlenecks to intramolecular energy flow in OCS. Therefore, the observation of the DH and NH hubs in the current work lends further support to the notion that lower dimensional phase space structures may act as traps and barriers for the classical dynamics in systems with high degrees of freedom. 
We now turn to a closer analysis of the nature of these dynamically 
significant regions of the web.

\subsection{Role of the dissociation hub}
\label{hubsrole}
The global phase space dynamics shown in Fig.~\ref{fig2} involves an ensemble of trajectories with constant actions but varying angles. In order to understand the role of the DH and NH hubs it is useful to distinguish between various dynamical classes of trajectories.
Specifically, the classification in terms of 
regular, chaotic but nondissociating, and chaotic dissociating trajectories
can be obtained from the computed nonlinear frequencies. 

\begin{center}
\begin{figure*}[t]
\includegraphics[width=0.9\textwidth]{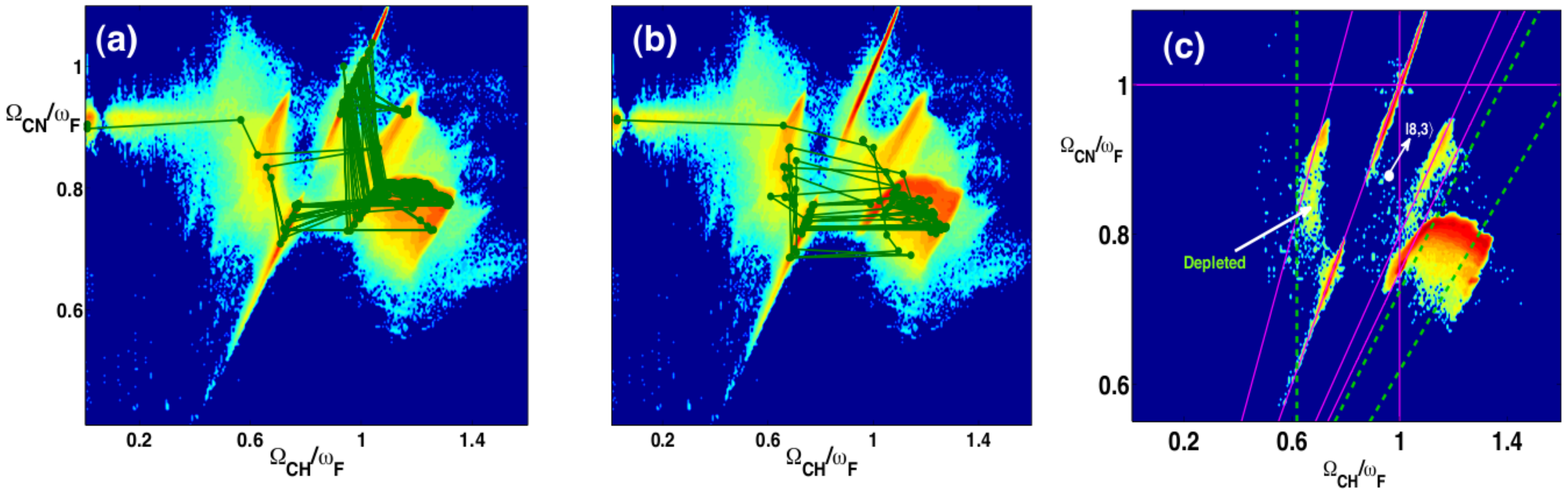}
\caption{Examples of different classes of dissociating chaotic trajectories and the importance of DH and NH regions. 
(a) A  trajectory shuttles between resonance zones and eventually dissociates
at $3300$ $T_{F}$. (b) A fast dissociating trajectory
which rapidly dissociates by $300$ $T_{F}$. Note that the trajectories
in (a) and (b) cross the DH region, indicated in Fig.~\ref{fig2}, and then dissociate.
(c) Frequency ratio space for  $|8,3\rangle$ constructed by
considering the trajectories that do not dissociate in presence of the laser field
until the final time of $5000$ $T_{F}$. Note that in (a) and (b) the points are plotted at intervals of $6.5$ $T_{F}$ and $2.5$ $T_{F}$ respectively for clarity.}
\label{fig3}
\end{figure*}
\end{center}

In Fig.~\ref{fig3}(a), a representative chaotic 
trajectory with a very large dissociation time is shown in the frequency ratio space.
It is evident from the figure that the trajectory jumps between the important
hubs and resonances in the web. However, interestingly, at a specific time the trajectory crosses the DH hub
and promptly dissociates. We observed this to be a feature that is common to several such trajectories which dissociate over long timescales.
In contrast, a rapidly dissociating chaotic trajectory shown in Fig.~\ref{fig3}(b), undergoes very different dynamics enroute to dissociation. On comparing with the dynamics in Fig.~\ref{fig3}(a), there is a marked decrease in the jumps between the $\Omega_{CH}:\Omega_{CN}=1:1$ region and the NH region. Nevertheless, prompt dissociation again seems to occur once the trajectory crosses the DH region at a specific time.
On the other hand, a regular trajectory (not shown here) remains
locked in  the $\Omega_{CH}:\Omega_{CN}=1:1$ resonance and a chaotic but non-dissociating trajectory
jumps between the $\Omega_{CH}:\Omega_{CN}=1:1$ resonance
and the noble hub
over the entire time duration. More importantly, the nondissociating trajectories, regular and chaotic, 
do not cross the specific DH region in the frequency ratio space.

Is crossing of the dissociation hub a generic feature of all the dissociating trajectories?
In order to answer this question, 
 in Fig.~\ref{fig3}(c) we show the frequency ratio space
from a data set containing only non-dissociating trajectories.
Interestingly, Fig.~\ref{fig3}(c) shows that the density in the vicinity of the
DH hub diminishes whereas the density around the NH region remains unchanged.
This clearly indicates that the high density at the dissociation hub in
Fig.~\ref{fig2} is a specific characteristic of dissociating trajectories. 
Therefore, the DH region is like a ``gateway" for dissociation since the frequency ratio space is also rather sparse to the left of the DH region.

\subsection{Role of the noble hub}

It is clear from the web shown in Fig.~\ref{fig2} that $1$:$1$ mode-mode resonance
and the noble hub consisting of several irrational frequency ratios, and their rational approximants in terms of high order
resonances, 
are dynamically the most visited regions of the web. 
The enhanced density in the NH region could be due to two main reasons.
Firstly, this might be because of frequent visits of every trajectory in the specific region of the frequency ratio space resulting in the observed enhanced density. 
Secondly,  the trajectories could be genuinely sticky and upon entering the region could be trapped for a considerable time. Such stickiness of individual trajectories can also lead to enhanced density. Establishing whether the NH region is sticky or not is crucial since the existence of sticky regions in phase space
imply long time dynamical correlations which, in turn, have important implications for 
the transport mechanism\cite{stickyall}. Note that Fig.~\ref{fig3}(c) already hints at the possibility of NH being a sticky region. In addition, our calculations (not shown here) of the mean square displacement of the action $J_{CH}$  
\begin{equation}
\langle \left(\Delta J_{CH}\right)^2 \rangle(\tau) \equiv \langle \left[J_{CH}(\tau)-J_{CH}(0)\right]^2 \rangle \sim \tau^{\beta}
\end{equation}
with the average being done over nondissociating trajectories, show $\beta < 1$ and hence subdiffusive behaviour.

\begin{center}
\begin{figure*}[t]
\includegraphics[width=0.9\textwidth]{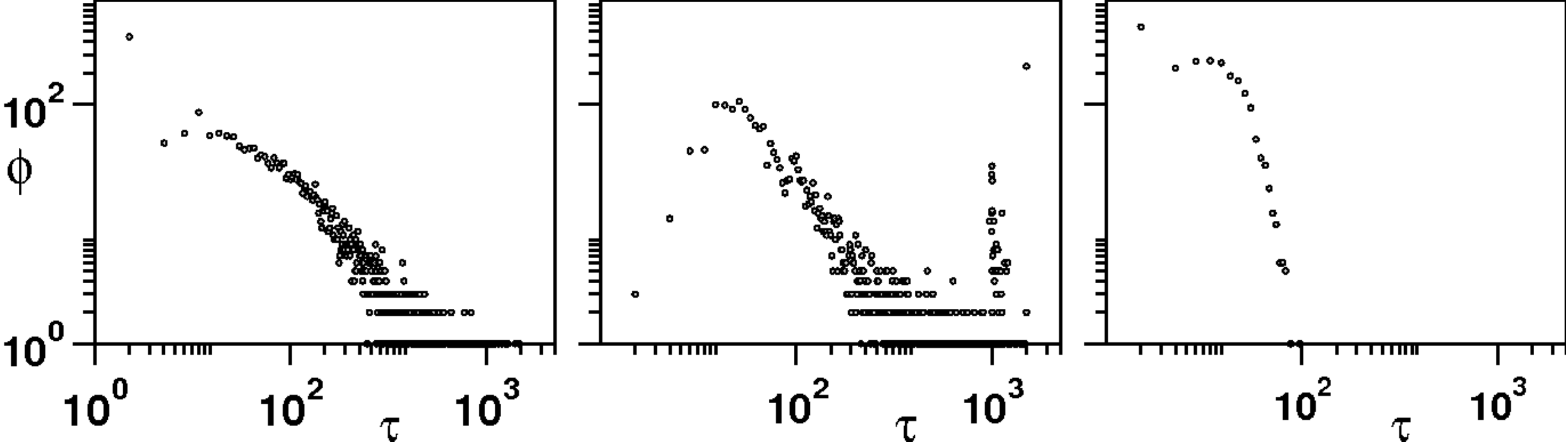}
\caption{Distribution $\phi$ of longest residence times in specific regions of the frequency ratio space corresponding to the initial state
$|8,3\rangle$. 
The panels from left to right show $\phi$ for the NH region, the $1$:$1$ mode-mode resonance region and the DH region
respectively.}
\label{fig5}
\end{figure*}
\end{center}

Further understanding into the nature of the NH region can be obtained by analysing the residence times of trajectories entering the region.
The approach employed to obtain the residence time distribution is as follows.
We define boxes in the high density regions of the
frequency ratio space as shown in Fig.~\ref{fig2}.
During the dynamics the time
at which the trajectory enters and then leaves this region are
denoted by $\tau_{1}$ and  $\tau_{2}$ respectively. The residence time $\tau_{r} \equiv \tau_{2}-\tau_{1}$ for one such event of
the trajectory residing in the NH region is noted.
There may be several events of such kind for a trajectory and out of all
such events we pick the longest $\tau_{r}$  in the region of
interest. We then construct the distribution of the longest $\tau_{r}$ for
the entire ensemble of trajectories for fixed initial actions (corresponding to the initial state of interest) and varying initial angles.

The distribution of the longest residence times in the
NH, $\Omega_{CH}=\Omega_{CN}$ resonance and DH regions are shown in Fig.~\ref{fig5}.
Clearly, the resonant and the NH regions do support trajectories with fairly long residence times. Therefore, the enhanced density in the NH region of the web in Fig.~\ref{fig3} arises due to extensive stickiness and supports the subdiffusive behaviour of $\langle (\Delta J_{CH})^2 \rangle(\tau)$ mentioned above. In comparison, the DH region is far less sticky and hence in agreement with it being interpreted as the dissociation gateway.
Although not shown here,  studies on other initial states do highlight similar regions in the frequency ratio space. The differences in the dissociation dynamics, however, arise from the relative extent of trapping in the various regions.
The results in this section imply that at least on the classical level it is not appropriate to describe the dissociation as being statistical. The three main hubs in the classical phase space play a crucial role in the dissociation mechanism.

As the key regions in the phase space have been identified, the next logical question is whether we can discover the mechanism of dissociation.  In order to answer this question we also need the precise mechanisms, in terms of appropriate phase space structures, by which  trajectories are transported from one region to the other. In particular, one needs to understand the transport near resonance junctions - a demanding\cite{wigbook} but important task for the future. Nevertheless, an  idea of the different mechanisms can be obtained from the dynamical frequency ratio space.
The jumps in Fig.~\ref{fig3}(a) happen to be along lines with negative slopes. On comparing to the static web in Fig.~\ref{fig1p} it seems that the ternary resonances are playing a crucial role. Although there are several high order mode-mode resonances in the vicinity of the NH region, it is useful to remember that the CH mode does not dissociate in the absence of the field. Clearly, significant excitations in the CN mode are required in order to cause extensive IVR and evidence for the same can be observed in Fig.~\ref{fig3} as visitations of the trajectory to regions with low values of $\Omega_{CN}$. As a rough estimate, a value of $\Omega_{CN}/\omega_{F} \approx 0.7$ corresponds to the unperturbed action value of $J_{CN} \approx 20$. Thus, the ternary resonances along with the $1$:$1$ resonance must be enhancing the IVR between the two modes, ultimately resulting in dissociation. However, Fig.~\ref{fig3} unequivocally shows that the  pathways of the short and long time
dissociation trajectories are different.

\section{Conclusions, challenges and future directions}
\label{conclusions}

In this work we have studied the classical dynamics of a system with $N = 2.5$ degrees of freedom to gain insights into the dissociation dynamics in presence of IVR.  Our work emphasizes the utility of time-frequency analysis in understanding the nature of phase space transport in such high dimensional phase spaces. 
In particular, we have constructed and identified regions
in the resonance web {\it i.e.,} in the full phase space, which control and regulate the classical dissociation dynamics of a model triatomic molecule. Three key regions in the web regulate the dynamics. Of these, 
 a region called as the dissociation hub acts as a gateway to dissociation and the region called as the noble hub,
corresponding to the overlap of
several high order mode-mode nonlinear resonances and
pairwise irrational frequency ratios, leads to extensive stickiness.
 The combined influence of the dissociation and noble hubs in the phase space results in the 
phase space transport being anomalous. Thus, despite being chaotic, certain trajectories survive
for very long time without undergoing dissociation. Moreover, the frequency ratio space dynamics clearly shows the existence of different dissociation pathways for prompt and late dissociating class of trajectories.

However, the present study represents only a start towards, what we believe, are important questions that need to be answered in order to understand and influence the mechanism of dissociation. We conclude by mentioning a few of these issues.
\begin{enumerate}
\item What is the precise mechanistic difference between the fast and slow dissociating trajectories? Some hints do emerge from the present work. However, a systematic study of the role of the various resonance junctions and the phase space structures which mediate transport between them is necessary to make further progress. In this regard it is imperative to construct\cite{arwebcomment} the full Arnold web and study the time evolution of the web.    

\item Is it possible to relate the notion of the dissociation hub as a gateway to the concept of transition states in driven systems\cite{kbjbpu07}?  This  requires one to determine the invariant manifolds in phase space using the technique of time-dependent normal forms and thereby distinguish between long time chaotic dissociating and nondissociating trajectories. Such a study would provide the much needed connection between the dynamics in the frequency ratio space and the powerful approach of transition state theory.  

\item Do the classical phase space structures and the resulting mechanisms survive quantization? 
This is an important issue and the studies reported here provide a firm baseline for establishing a clear
classical-quantum correspondence perspective into the role of IVR in quantum control.
Preliminary computations of the quantum dissociation probabilities
suggest that changes in the phase space dynamics caused by external driving fields, especially
stickiness, do influence the quantum transport as well. In particular, there are indications that the various hubs are more restrictive in the quantum case.       

\item  Given that the key
phase space structures regulating the dissociation dynamics have been identified, is it indeed possible to design control schemes by interfering with the
specific phase space structures? Such a local phase space approach towards control has been explored recently for
systems with lower degrees of freedom\cite{locpt1,locpt2}. It remains to be seen if the insights gained from the current work can be utilized for controlling the dissociation dynamics by locally rebuilding appropriate phase space barriers.
Furthermore, we hope that understanding the phase space dynamics of such higher dimensional
systems might provide a useful way of gaining insights into the nature of the  
fields coming out of an optimal control theory calculation. A recent work\cite{kk11} addresses this issue in the context of a driven one degree of freedom system and  the work presented in this paper is a first step towards generalizing to driven systems with higher degrees of freedom.

\item One aspect of the problem that we have been silent about is that of the phenomenon of Arnold diffusion. In principle, the system considered here can undergo Arnold diffusion. However, in practice, the timescale for Arnold diffusion might be too long compared to the dissociation timescale and hence play no role in the mechanism of slow dissociating trajectories. Nevertheless, from a fundamental perspective, the current system provides another model, apart from the driven coupled quartic oscillator model\cite{chirikov} considered by Chirikov in his seminal work, for studying Arnold diffusion and its quantum manifestations\cite{ardifquantum1,ardifquantum2,ardifquantum3}.
\end{enumerate}

\section*{Acknowledgments}

This work was supported by a fellowship to AS from 
Indian Institute of Technology, Kanpur, India. We would also like to thank Amber Jain for useful discussions during early stages of this work.



\begin{thebibliography}{}

\bibitem{al77}{R. V. Ambartsumian and V. S. Letokhov, {\em {Chemical and Biochemical Applications of
Lasers}} (C. B. Moore, Ed.), {\bf 3}, (Academic Press, New York, 1977).}

\bibitem{l77}{V. S. Letokhov, {\em {Phys. Today}}, {\bf 30}, 1977,p. 23.}

\bibitem{by78}{N. Bloembergen and E. Yablonovitch, {\em {Phys.
Today}}, {\bf 31}, 1978, p. 23.}

\bibitem{z80}{A. H. Zewail, {\em {Phys. Today}}, {\bf 33}, 1980, p. 27.}

\bibitem{jlr81}{J. Jortner, R. D. Levine and S. A. Rice, Adv. Chem. Phys. {\bf 47}, (Wiley, New York, 1981).}

\bibitem{or79}{I. Oref and B. S. Rabinovitch, Acc. Chem. Res. {\bf 12}, 166 (1979).}

\bibitem{zb88}{A. H. Zewail and N. Bloembergen, J. Phys. Chem. {\bf 88}, 5459 (1984).}

\bibitem{bsbook}{M. Shapiro and P. Brumer, {\em Principles of
the Quantum Control of Molecular Processes} (Wiley, New York, 2003).}

\bibitem{gr97}{R. J. Gordon and S. A. Rice, Annu. Rev. Phys. Chem. {\bf 48}, 601 (1997).}

\bibitem{rz00}{H. Rabitz and W. Zhu, Acc. Chem. Res. {\bf 33}, 572 (2000).}

\bibitem{exivr1}{J. C. Keske, D. A. McWhorter and  B. H. Pate, Int. Rev. Phys. Chem. {\bf 19}, 363 (2000).}

\bibitem{exivr2}{B. C. Dian, G. G. Brown, K. O. Douglass, F. S. Rees, J. E. Johns, P. Nair,
R. D. Suenram and B. H. Pate, Proc. Natl. Acad. Sci.  {\bf 105}, 12696 (2008).}

\bibitem{exivr3}{E. W. G. Diau, J. L. Herek, Z. H. Kim and A. H. Zewail, Science {\bf 279}, 847 (1998).}

\bibitem{thivr1}{U. Lourderaj and W. L. Hase, J. Phys. Chem. A {\bf 113}, 2236 (2009).}

\bibitem{thivr2}{S. C. Farantos, R. Schinke, H. Guo and M. Joyeux, Chem. Rev. {\bf 109}, 4248 (2009).}

\bibitem{thivr3}{D. M. Leitner, Int. J. Quan. Chem. {\bf 75}, 523 (1999).}

\bibitem{stsp0}{M. Gruebele, Adv. Chem. Phys. {\bf 114}, 193 (2000).}

\bibitem{stsp1}{M. Gruebele and P. G. Wolynes, Acc. Chem. Res. {\bf 37}, 261 (2004).}

\bibitem{stsp2}{D. M. Leitner, Adv. Chem. Phys. {\bf 130B}, 205 (2005).}

\bibitem{sw93}{S. Schoﬁeld and P. G. Wolynes, J. Chem. Phys. {\bf 98}, 1123 (1993).}

\bibitem{mg98}{M. Gruebele, Proc. Nat. Acad. Sci. {\bf 95}, 5965 (1998).}

\bibitem{ivrcontrol1}{M. Sugawara, Adv. Phys. Chem. {\bf 2011}, 584082 (2011).}

\bibitem{ivrcontrol2}{M. Gruebele, Theor. Chem. Acc. {\bf 109}, 53 (2003).}

\bibitem{ivrcontrol3}{D. Gerbasi, A. S. Sanz, P. S. Christopher, M. Shapiro and P. Brumer, J. Chem. Phys. {\bf 126}, 124307 (2007).}

\bibitem{onfzr2001}{Y. Ohtsuki, K. Nakagami, Y. Fujimura, W. Zhu and H. Rabitz, J. Chem. Phys. {\bf 114}, 8867 (2001).}

\bibitem{bgm2011}{L. Blancafort, F. Gatti and H. D. Meyer, J. Chem. Phys. {\bf 135}, 134303 (2011).}

\bibitem{swr88}{S. Shi, A. Woody and H. Rabitz,  J. Chem. Phys. {\bf 88}, 6870 (1988).}

\bibitem{dfs}{M. P. A. Branderhorst, P. Londero, P. Wasylczyk, C. Brif, R. L. Kosut, H. Rabitz and I. A. Walmsley, Science {\bf 320}, 638 (2008).}

\bibitem{drivatom}{P. Koch and K. Van Leeuwen, Phys. Rep. {\bf 255}, 289 (1995).}

\bibitem{mesoqc}{H. Schomerus and P. Jacquod, J. Phys. A: Math. Gen. {\bf 38}, 10663 (2005).}

\bibitem{bec}{R. Gati and M. K. Oberthaler, J. Phys. B: At. Mol. Opt. Phys. {\bf 40}, R61 (2007).}

\bibitem{cl80}{S. R. Channon and J. Lebowitz,  Ann. N. Y. Acad. Sci. {\bf 357}, 108 (1980).}

\bibitem{cantorirefs}{R. S. MacKay, J. D. Meiss and I. C. Percival, Physica D {\bf 13}, 55 (1984).}

\bibitem{psbifur}{M. E. Kellman and V. Tyng, Acc. Chem. Res. {\bf 40}, 243 (2007).}

\bibitem{ivr2drev1}{T. Uzer, Phys. Rep. {\bf 199}, 73 (1991).}

\bibitem{ivr2drev2}{G. S. Ezra, Adv. Class. Traj. Meth. {\bf 3}, 35 (1998).}

\bibitem{ftl}{C. Froeschl\'{e}, M. Guzzo and E. Lega, Science {\bf 289}, 2108 (2000).}

\bibitem{megno}{P. M. Cincotta, C. M. Giordano and C. Sim\'{o},  Physica D {\bf 182}, 151 (2003).}

\bibitem{fmi}{B. Cordani, Physica D {\bf 237}, 2797 (2008).}

\bibitem{stickyall}{ A. D. Perry and S. Wiggins, Physica D {\bf 71}, 102 (1994).} 

\bibitem{sticky0}{Y. C. Lai, M. Ding, C. Grebogi and R. Bl\"{u}mel, Phys. Rev. A {\bf 46}, 4661 (1992).}

\bibitem{sticky1}{G. M. Zaslavsky, Phys. Rep. {\bf 371}, 461 (2002).}

\bibitem{sticky2}{E. G. Altmann, A. E. Motter and H. Kantz, Phys. Rev. E {\bf 73}, 026207 (2006).}

\bibitem{sticky3}{H. Tanaka and A. Shudo, Phys. Rev. E {\bf 74}, 036211 (2006).}

\bibitem{sticky4}{R. Venegeroles, Phys. Rev. Lett. {\bf 102}, 064101 (2009).}

\bibitem{sticky5}{Y. S. Sun and L. Y. Zhou, Celest. Mech. Dyn. Astron. {\bf 103}, 119 (2009).}

\bibitem{sticky6}{K. Tsiganis, A. Anastasiadis and H. Varvoglis, Chaos, Solitons and Fractals {\bf 11}, 2281 (2000).}

\bibitem{tsrev1}{H. Waalkens, R. Schubert and S. Wiggins, Nonlinearity {\bf 21}, R1 (2008).}

\bibitem{tsrev2}{T. Uzer, C. Jaff\'{e}, J. Palacian, P. Yanguas and S. Wiggins, Nonlinearity {\bf 15}, 957 (2002).}

\bibitem{tsrev3}{D. M. Leitner, Y. Matsunaga, C. B. Li, T. Komatsuzaki, A. Shojiguchi and M. Toda, Adv. Chem. Phys. {\bf 145}, 83 (2011).}

\bibitem{kbjbpu07}{S. Kawai, A. D. Bandrauk, C. Jaff\'{e}, T. Bartsch, J. Palaci\'{a}n and T. Uzer, J. Chem. Phys. {\bf 126}, 164306 (2007).}

\bibitem{webref1}{A. Shojiguchi, C. B. Li, T. Komatsuzaki and M. Toda, Phys. Rev. E {\bf 76}, 056205 (2007).}

\bibitem{webref2}{A. Semparithi and  S. Keshavamurthy, J. Chem. Phys. {\bf 125}, 141101 (2006).}

\bibitem{webref3}{P. Manikandan, A. Semparithi and S. Keshavamurthy, J. Phys. Chem. A {\bf 113}, 1717 (2009).}

\bibitem{cb91}{S. Chelkowski and A. D. Bandrauk, Chem. Phys. Lett. {\bf 186}, 264 (1991).}

\bibitem{bl04}{R. Brezina and W. K. Liu,  J. Phys. Chem. A {\bf 108}, 8852 (2004).}

\bibitem{hobi02}{ R. Hasbani, B. Ostoji´c, P. R. Bunker and M. Yu. Ivanov, J. Chem. Phys. {\bf 116}, 10636 (2002).}

\bibitem{brr95}{J. Botina, H. Rabitz and N. Rahman, J. Chem. Phys. {\bf 102}, 226 (1995).}

\bibitem{mde87}{C. C. Martens, M. J. Davis and G. S. Ezra, Chem. Phys. Lett. {\bf 142}, 519 (1987).} 

\bibitem{tw61}{E. Thiele and D. J. Wilson, J. Chem. Phys. {\bf 35}, 1256 (1961).}

\bibitem{oxrice76}{D. W. Oxtoby and S. A. Rice, J. Chem. Phys. {\bf 65}, 1676 (1976).}

\bibitem{gogmil88}{M. E. Goggin and P. W. Milonni, Phys. Rev. A {\bf 37}, 796 (1988).}

\bibitem{gmr05}{J. Gong, A. Ma and S. A. Rice, J. Chem. Phys. {\bf 122}, 144311 (2005).}

\bibitem{H40}{K. Husimi, Proc. Math. Soc. Jpn. {\bf 22}, 264 (1940).}

\bibitem{mw82}{D.L. Martin and R.E. Wyatt, Chem. Phys. {\bf 64}, 203 (1982).}

\bibitem{chiricomment}{This result can be anticipated by performing a Chirikov overlap analysis on the driven system following Goggin and Milonni\cite{gogmil88}. Estimates show that, within a linearized dipole approximation, the threshold field strength $\lambda_{F} > 8.7 \times 10^{-2}$ au for overlap between the $\omega_{F}:\Omega_{CH} = 1:1$ and $\omega_{F}:\Omega_{CH} = 2:1$ nonlinear resonances.}

\bibitem{arwebcomment}{Note that the results reported here are not quite the full Arnold web and can be considered as a ``coarse" web. The full Arnold web entails much higher computational efforts\cite{ftl,megno} and also reveals the stability of the various regions in the web. Although in this preliminary study we do not undertake such a task, it is interesting that the wavelet approach itself can be utilized to map out the full web by the recently proposed method of frequency modulation indicator\cite{fmi}.}

\bibitem{l93}{J. Laskar, Physica D {\bf 67}, 257 (1993).}

\bibitem{aw01}{L. V. Vela-Arevalo and S. Wiggins, Int. J. Bifur. Chaos. {\bf 11}, 1359 (2001).}

\bibitem{pcu2009}{R. Paskauskas, C. Chandre and T. Uzer, J. Chem. Phys. {\bf 130}, 164105 (2009).}

\bibitem{wigbook}{S. Wiggins, {\em Chaotic Transport in Dynamical Systems} (Springer-Verlag, New York, 1992).}

\bibitem{locpt1}{S. Huang, C. Chandre and T. Uzer, Phys. Rev. A {\bf 74}, 053408 (2006).}

\bibitem{locpt2}{A. Sethi and S. Keshavamurthy, Phys. Rev. A {\bf 79}, 033416 (2009).}

\bibitem{kk11}{S. Kawai and T. Komatsuzaki, J. Chem. Phys. {\bf 134}, 024317 (2011).}

\bibitem{chirikov}{B. V. Chirikov, Phys. Rep. {\bf 52}, 263 (1979).}

\bibitem{ardifquantum1}{D. M. Leitner and P. G. Wolynes, Phys. Rev. Lett. {\bf 79}, 55 (1997).}

\bibitem{ardifquantum2}{V. Ya. Demikhovskii, F. M. Izrailev and A. I. Malyshev, Phys. Rev. E  {\bf 66}, 036211 (2002).}

\bibitem{ardifquantum3}{A. I. Malyshev and L. A. Chizova, J. Expt. Theor. Phys. {\bf 110}, 837 (2010).}

\end{thebibliography}
\end{document}